\documentclass[aps,prl,twocolumn,superscriptaddress,notitlepage,nofootinbib]{revtex4-1}
\usepackage{amsmath}
\usepackage{epsfig}
\usepackage{color,soul}
\usepackage{multirow}

\def\bea#1\eea{\begin{align}#1\end{align}}

\newcommand{\nn}{\nonumber\\}

\newcommand{\bef}{\begin{figure}[h!tb]\centering}
\newcommand{\eef}{\end{figure}}
\newcommand{\as}{\alpha_s}

\renewenvironment{widetext@grid}{%
  \par\ignorespaces
  \setbox\widetext@top\vbox{%
   \vskip15\p@
   \hb@xt@\hsize{%
    \leaders\hrule\hfil
    \vrule\@height6\p@
   }%
   \vskip6\p@
  }%
  \setbox\widetext@bot\hb@xt@\hsize{%
    \vrule\@depth6\p@
    \leaders\hrule\hfil
  }%
  \onecolumngrid
  \let\set@footnotewidth\set@footnotewidth@ii
}{%
  \par
  \twocolumngrid\global\@ignoretrue
  \@endpetrue
}%

\begin{document}
\title{Threshold and jet radius joint resummation for single-inclusive jet production}

\begin{flushright}
DESY 17-119 \\
\end{flushright}

\author{Xiaohui Liu}
\email{xiliu@bnu.edu.cn}
\affiliation{Center of Advanced Quantum Studies, Department of Physics, Beijing Normal University, Beijing 100875, China}

\author{Sven-Olaf Moch}
\email{sven-olaf.moch@desy.de}
\affiliation{II. Institut f\"ur Theoretische Physik, Universit\"at Hamburg, Luruper Chaussee 149, D-22761 Hamburg, Germany} 

\author{Felix Ringer}
\email{fmringer@lbl.gov}
\affiliation{Nuclear Science Division, Lawrence Berkeley National Laboratory, Berkeley, California 94720, USA}
                   
\date{\today}         

\begin{abstract}
We present the first threshold and jet radius jointly resummed cross section for single-inclusive hadronic jet production. We work at next-to-leading logarithmic accuracy and our framework allows for a systematic extension beyond the currently achieved precision. Longstanding numerical issues are overcome by performing the resummation directly in momentum space within Soft Collinear Effective Theory. We present the first numerical results for the LHC and observe an improved description of the available data. Our results are of immediate relevance for LHC precision phenomenology including the extraction of parton distribution functions and the QCD strong coupling constant.
\end{abstract}

\date{\today}

\maketitle
{\it Introduction.} The inclusive production of jets plays a crucial role at the LHC and the corresponding cross section has been measured with great accuracy by ALICE, ATLAS and CMS~\cite{Abelev:2013fn,Chatrchyan:2014gia,Khachatryan:2016jfl,Aaboud:2017dvo,Khachatryan:2016wdh,ATLAS-CONF-2017-048}. From the theoretical point of view, inclusive jet production constitutes a benchmark process that is used to determine universal non-perturbative quantities like parton distribution functions (PDFs) and the QCD strong coupling constant $\as$. In this sense, a very good understanding of the relevant QCD dynamics for inclusive jet production at the LHC is crucial as it will impact the comparisons between theory and data for other processes as well. Furthermore, high transverse momentum jets are promising observables for the search of physics beyond the standard model. 

In order to match the achieved experimental precision for the process $pp\to\text{jet}+X$, ongoing theory efforts have recently succeeded in calculating the fully differential cross section at next-to-next-to leading order (NNLO)~\cite{Currie:2016bfm,Currie:2017ctp}. The results were presented for all partonic processes in the leading-color approximation for the $\alpha_s^2$ coefficient. While the completion of the NNLO results marks a new milestone for high precision QCD calculations, there are, nevertheless, remaining theoretical uncertainties. 
Recent comparisons of the NNLO predictions with the ATLAS measurements suggest that even at NNLO the results still heavily rely on the scale choice~\cite{ATLAS-CONF-2017-048}. Slightly different scale choices can lead to quite different NNLO predictions which indicates large higher-order perturbative corrections as well as an underestimation of the QCD scale dependence as pointed out in~\cite{Dasgupta:2016bnd,Kang:2016mcy}. From a practical point of view, any information beyond fixed NNLO accuracy can only be accessed by using resummation techniques, where dominant classes of logarithms are summed up to all orders in the strong coupling constant. In this work, we focus specifically on the joint resummation of the following two numerically important classes relevant for the current experimental kinematics: threshold logarithms and logarithms in the jet-size parameter $R$. 

The importance of resumming single logarithms in the jet-size parameter $\alpha_s^n\ln^n R$ was addressed in~\cite{Dasgupta:2014yra,Kang:2016mcy,Dai:2016hzf,Dai:2017dpc}. The so-called threshold logarithms arise near the exclusive phase space boundary, where the production of the signal-jet just becomes possible. At threshold, the invariant mass $\sqrt{s_4}$ of the unobserved partonic system recoiling against the signal-jet vanishes. Note that the signal-jet retains a finite invariant mass at threshold allowing for radiation inside the jet cone~\cite{Mukherjee:2012uz,deFlorian:2013qia}. The cancelation of infrared divergences leaves behind logarithms of the form $\as^n(\ln^k (z)/z)_+$, with $k\leq 2n-1$, and $z=s_4/s$, where $s$ is the partonic center-of-mass (CM) energy. In the threshold limit as $z\to 0$, these terms become large and need to be resummed to all orders so as to obtain reliable perturbative results. In~\cite{deFlorian:2013qia}, it was shown that threshold logarithms dominate indeed over a wide range of the jet-$p_T$ even far away from the hadronic threshold due to the steeply falling parton luminosity functions. 

Even though the threshold resummed cross section for hadronically produced jets was addressed before~\cite{Kidonakis:1998bk,Kidonakis:1998nf}, it has so far eluded a numerical evaluation. Traditionally, threshold resummation is derived in Mellin moment space~\cite{Sterman:1986aj,Catani:1989ne,Catani:1996yz} and was applied to the rapidity integrated inclusive jet cross section in~\cite{deFlorian:2007fv} at next-to-leading logarithmic (NLL) accuracy. However, in order to allow for a meaningful comparison to the available data, the complete kinematics of the jet have to be taken into account. The traditional methods failed to apply in this case so far. The reasons are twofold and can be traced back to the factorization structure of the resummed cross section and the specific properties of the Mellin transformation. Instead, only fixed-order (FO) expansions of the threshold resummed cross section are currently available in the literature~\cite{Kidonakis:2000gi,Kumar:2013hia,deFlorian:2013qia}. Note that these problems do not necessarily occur for observables with identified final state hadrons~\cite{deFlorian:2013taa,Catani:2013vaa,Hinderer:2014qta,Uebler:2015ria}.

In this work, we present for the first time the results for the threshold and small-$R$ jointly resummed inclusive jet cross section in proton-proton collisions. The shortcomings of the traditional approaches to threshold resummation are overcome by making use of techniques developed in the context of Soft Collinear Effective Theory (SCET)~\cite{Bauer:2000ew,Bauer:2000yr,Bauer:2001ct,Bauer:2001yt,Beneke:2002ph}, which 
allows for the resummation to be carried out directly in momentum space~\cite{Becher:2006nr}. 
Since there are no numerical results available for the threshold resummed inclusive jet cross section using traditional methods, it is here, where the 
SCET approach exhibits its full potential. In addition, our framework allows for a systematic extension to next-to-next-to-leading-logarithmic (NNLL) accuracy or beyond for the resummation of both threshold and the small-$R$ logarithms, which we briefly discuss below and address in detail in a future publication.

{\it Theoretical framework.}  
The double differential cross section for the process $pp\to\text{jet}+X$ can be written as 
\bea
\frac{ p_T^2 \mathrm{d}^2 \sigma    }{   \mathrm{d}p_T^2 \mathrm{d} \eta } 
& =  \sum_{i_1i_2} 
\int_0^{V(1-W)}  \mathrm{d} z 
\int_{\frac{VW}{1-z}}^{1- \frac{1-V}{1-z}}  \mathrm{d}v \,
x_1^2\,  f_{i_1}(x_1) \, x_2^2 \,  f_{i_2}(x_2)   &  \nn 
 & \times \frac{\mathrm{d}^2 \hat{\sigma}_{i_1i_2}}{ \mathrm{d} v \, \mathrm{d} z } (v,z,p_T,R)   \,,  &
\eea
where $p_T$ and $\eta$ are the transverse momentum and rapidity of the signal-jet, respectively, 
and we have $V = 1-  p_T e^{-\eta}/\sqrt{S}$, $VW = p_T e^{\eta}/\sqrt{S} $ and the hadronic CM energy is denoted by $\sqrt{S}$. 
The sum runs over all partonic channels initiating the process whose cross sections are given by $\hat{\sigma}_{i_1i_2}$. 
Besides depending on $p_T$, the partonic cross sections $\hat{\sigma}_{i_1i_2}$ are functions of the partonic kinematic variables 
$s = x_1 x_2 S$, 
$v = u/(u+t) $ and $z$. Here we have introduced $t = (p_1 - p_3)^2$ and $u = (p_2 - p_3)^2$, where $p_{1,2}$ are the momenta of the two incoming partons and $p_3$ is the momentum of the parton initiating the signal-jet. The PDFs are denoted by $f_i$ evaluated at the momentum
fractions $x_1 = VW/v/(1-z)$ and $x_2 = (1-V)/(1-v)/(1-z) $.  

In the small-$R$ and $z\to0$ threshold limit, the partonic cross sections can be further factorized as
\bea\label{eq:fac}
& \frac{\mathrm{d}^2 \hat{\sigma}_{i_1i_2}}{ \mathrm{d} v \, \mathrm{d} z }   = 
 s \int \mathrm{d} s_X \, \mathrm{d}s_c \mathrm{d}s_G \, \delta(z s -s_X -s_G - s_c)  & \nn
&  \hspace{-1.5mm} 
\times {\rm Tr} \left[   { \bf H}_{i_1i_2}(v,p_T\,, \mu_h\,, \mu) \,  {\bf S}_G (s_G\,,\mu_{sG}\,, \mu) \right] 
J _X(s_X\,, \mu_{X} \,, \mu ) 
& \nn 
& \hspace{-1.5mm} \times  \sum_m 
{\rm Tr}\left[  J_{m}(p_T R \,,\mu_J\,, \mu) \otimes_{\Omega} S_{c,m}(s_c R\,, \mu_{sc} \,,  \mu )  \right]  \,, &
\eea
where the traces are taken in color space. The sum runs over all collinear splittings and `$\otimes_{\Omega}$\!' denotes the associated angular integrals~\cite{Becher:2015hka}. Here we have assumed that the jet is constructed using the anti-$k_T$ algorithm~\cite{Cacciari:2008gp}, $z\sim R$, and we allow for a finite mass of the signal-jet. The factorization formula is established within the framework of SCET,  where
 ${\bf H}_{i_1i_2}$ are the hard functions for $2 \to 2$ scattering, which are known to 2-loops~\cite{Broggio:2014hoa}. The inclusive jet function $J_X(s_X)$ depends on the invariant mass $s_X$ of the recoiling collimated radiation, and it is also known 
to order $ \alpha_s^2$~\cite{Becher:2006qw,Becher:2010pd}. The global soft function ${\bf S}_{G}$ takes into account wide-angle soft radiation which cannot resolve
the small jet radius $R$. At NLO, the bare global soft function ${\bf S}_G$ is found to be 
\bea
{\bf S}_G^{(1)} =  \frac{ \alpha_s}{ \pi \, \epsilon  }  \frac{e^{\epsilon \gamma_E }}{\Gamma(1-\epsilon) }     \sum_{i\ne j \ne 4 }  \!\!\!\!  {\bf T}_i \! \cdot \! {\bf T}_j   
 \frac{n_{ij} }{\mu_{sG} } \left(  \frac{s_G n_{ij}}{  \mu_{sG}}  \right)^{-1- 2\epsilon}  \,, 
\eea
with $n_{ij} = \sqrt{s_{ij}/s_{i4}/s_{j4} } $ and $s_{ij} = 2 p_i \cdot p_j$.   After
performing the renormalization in the $ \overline{ {\rm MS} }$
scheme, the NLO global soft function can be obtained as well as the anomalous dimension governing its renormalization group (RG) evolution. The signal-jet function $J(p_T\, R)$ and the soft-collinear (``coft'') function $S_c(s_c R)$~\cite{Becher:2015hka, Chien:2015cka} account for the energetic radiation inside the jet and the soft radiation near the jet boundary, respectively. Due to the fact that the soft-collinear radiation can resolve the splitting details of the collinear radiation inside the signal-jet, one has to perform an infinite sum over the collinear splitting history inside the jet and keep the angular correlations between the jet and soft-collinear radiation which account for the non-global logarithms (NGLs)~\cite{Dasgupta:2001sh}, as addressed in~\cite{Caron-Huot:2015bja, Larkoski:2015zka, Becher:2015hka, Larkoski:2016zzc}. We note that the signal-jet and the soft-collinear functions can be viewed as the threshold limit of the semi-inclusive jet function~\cite{Kang:2016mcy, Dai:2017dpc}. 
If we ignore the NGLs, which usually show their major effects in the deep-resummation region~\cite{Jouttenus:2013hs,Dai:2017dpc} and have a relatively small phenomenological impact for more inclusive cross sections~\cite{Jouttenus:2013hs,Liu:2012sz, Dai:2017dpc}, the infinite sum and the angular correlation structure can be approximated by a product of the jet and soft-collinear functions. The NLO jet function can be 
extracted from~\cite{Liu:2012sz} and for the NLO bare soft-collinear function, we find  (see also~\cite{Dai:2017dpc})
\bea
S_c^{(1)} ={\bf T}_3^2\,  \frac{\alpha_s }{\pi \, \epsilon}  \, 
\frac{e^{\epsilon \gamma_E} }{\Gamma(1-\epsilon) }
\frac{ p_T R }{  s \mu_{sc} }   
\left( \frac{s_c \, p_T R }{ s\, \mu_{sc} }\right)^{-1-2\epsilon}    \,,
\eea
from which the renormalized soft-collinear function and its anomalous dimension can be readily obtained.

In order to evaluate the cross section in Eq.~(\ref{eq:fac}), all functions are evolved from their natural scales $\mu_i$ to the scale $\mu$ according their RG equations which leads to the resummation of the large logarithms. Here, we do not elaborate on the solution of the various RG equations as this has been studied extensively in the literature, see for instance~\cite{Becher:2006nr}. With all currently available ingredients, Eq.~(\ref{eq:fac}) allows us to achieve the NLL resummation for hadronic single-inclusive jet production. In order to go beyond NLL accuracy, the relevant anomalous dimensions need to be extracted from explicit 2-loop calculations which are in principle within reach. The 2-loop hard and inclusive jet functions are both known and
  the 2-loop global soft and the soft-collinear functions can be obtained
  following~\cite{Becher:2012za} and~\cite{Kelley:2011ng,
    Boughezal:2015eha}. The angular correlation between the jet and
  soft-collinear function arises from integrating over the single soft limit of
  the $1\to 3$ splitting functions~\cite{Becher:2015hka} with appropriate
  phase space restrictions.   
The 2-loop signal jet function can be calculated at least numerically following the strategies of~\cite{Ritzmann:2014mka, Boughezal:2015eha}.
Alternatively, the anomalous dimensions, hence all $\ln R$ terms of the signal
jet function, can be extracted at 2-loop using consistency relations. The
remaining constant terms of the signal jet function can be determined
numerically using EVENT2~\cite{Catani:1996vz}. Here we are utilizing the fact
that the same jet function will appear for jet production at threshold in
$e^+e^-$ annihilation where a similar joint resummation formalism applies.

%
\begin{figure}[t]
\begin{center}
\vspace*{0cm}
\hspace*{0mm}
\epsfig{figure=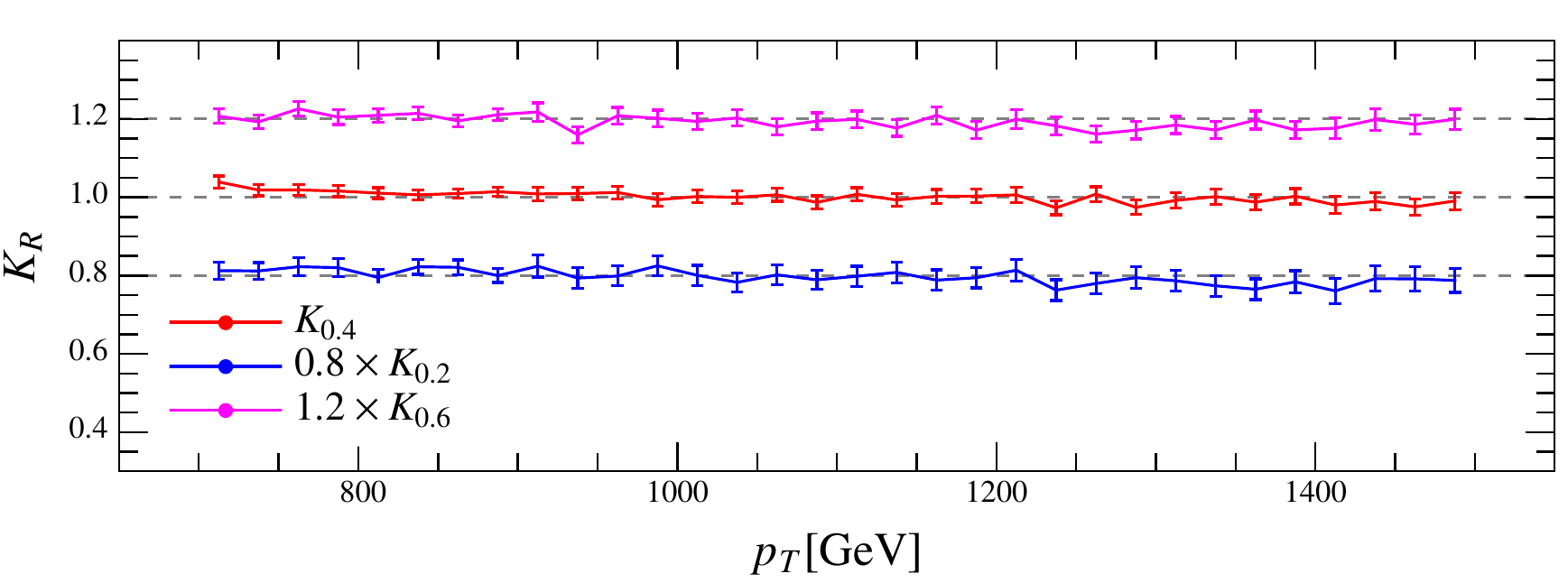,width=.47\textwidth,angle=0}
\end{center}
\vspace*{-.6cm}
\caption{Ratios $K_R$ of the ${\rm NLO}_{\rm sin}$ result which is obtained by expanding Eq.~(\ref{eq:fac}) and the full NLO QCD result for different values of $R$ as a function of the jet-$p_T$, for $1.5<|\eta|<2$ at $\sqrt{S}=13$~TeV. The error bars show the numerical uncertainty. 
\label{fig1}  }
\vspace*{-0.8cm}
\end{figure}
%
  
{\it Phenomenology.} 
To proceed, we first validate the factorization formalism by comparing the predictions of Eq.~(\ref{eq:fac}) expanded to NLO, denoted by ${\rm NLO}_{\rm sin}$ in the following, with the full NLO QCD calculation in the threshold region.  
Since Eq.~(\ref{eq:fac}) is derived in the strict threshold limit, the scale choice related to the jet-$p_T$ can only be the leading-jet transverse momentum $p_T^{\rm max}$, since no jets in the event can be harder than the signal-jet in this limit. Therefore, when comparing the two results, we choose the renormalization and factorization scales as $\mu = \mu_F = \mu_R = p_T^{\rm max}$ for the full NLO QCD calculation instead of using the so-called individual jet-$p_T$ which probes a softer scale than $p_T^{\rm max}$. We use the MMHT2014nlo PDF set of~\cite{Harland-Lang:2014zoa} and focus on  $\sqrt{S} = 13$~TeV.  
To enforce the threshold limit, we demand that $p_T > 700~{\rm GeV} $ and $1.5 < |\eta| <2 $. Fig.~\ref{fig1} displays the ratios $K_R$ of  the ${\rm NLO}_{\rm sin}$ result to the full NLO QCD calculation~\cite{Gao:2012he} for $R = 0.2$, $0.4$ and $0.6$. We find very good agreement between these two calculations for all choices of $R$ which validates our factorization theorem. 

We further separate the ${\rm NLO}_{\rm sin}$ result into a ``virtual'' $\delta(z)$ term and
the logarithmic terms $(\ln^k(z)/z)_+$ with $k = 0,1$. We observe that the ``virtual" term gives a large 
positive correction. The net logarithmic contribution decreases the cross section, where the $(\ln(z)/z)_+$ term is positive whereas the $(1/z)_+$ term is negative and large due to its coefficient in the kinematic regime under study.

Now we turn to the phenomenology at the LHC. We match the NLL resummed results with the full NLO calculation using
\bea\label{eq:match}
\mathrm{d} \sigma = \mathrm{d} \sigma_{\rm NLL} - \mathrm{d} \sigma_{{\rm NLO}_{\rm sin}} + \mathrm{d} \sigma_{\rm NLO} \,,
\eea
and we set $\mu = p_T^{\rm max}$ for the reasons discussed above. We make the central scale choices $\mu_h = p_T$ 
and $\mu_J = p_T R$ for the hard and the signal-jet functions, respectively. The naive scale for the recoiling jet function is of order $\mu_X  \sim  \kappa \, \sqrt{s} \left(1 - \frac{2p_T}{\sqrt{S}} \right)$ with $\kappa \sim 1$. However, due to the steeply falling shape of the luminosity function, $\kappa$ can deviate from $1$ and approach a smaller value. We determine $\kappa$ dynamically following~\cite{Becher:2009th} and we set $\mu_X = \kappa \times 2 p_T  \left(1 - \frac{2p_T}{\sqrt{S}} \right) $ with $\kappa = 1/2$. The other scales are determined in the seesaw way: $\mu_{sG} = \mu_X^2 / \mu_h $ and $\mu_{sc} = \mu_J \times  \mu_{sG} / \mu_h$. Our uncertainty estimates are obtained by varying $\mu$, $\mu_h$ and $\mu_J$ independently while keeping the seesaw relations for $\mu_X$, $\mu_{sG}$ and $\mu_{sc}$. For all scales we consider variations by a factor of $2$ around their central values and the final scale uncertainty is obtained by taking the envelope.
%
\begin{figure*}[t]
\begin{center}
\vspace*{-2cm}
\hspace*{-5mm}
\epsfig{figure=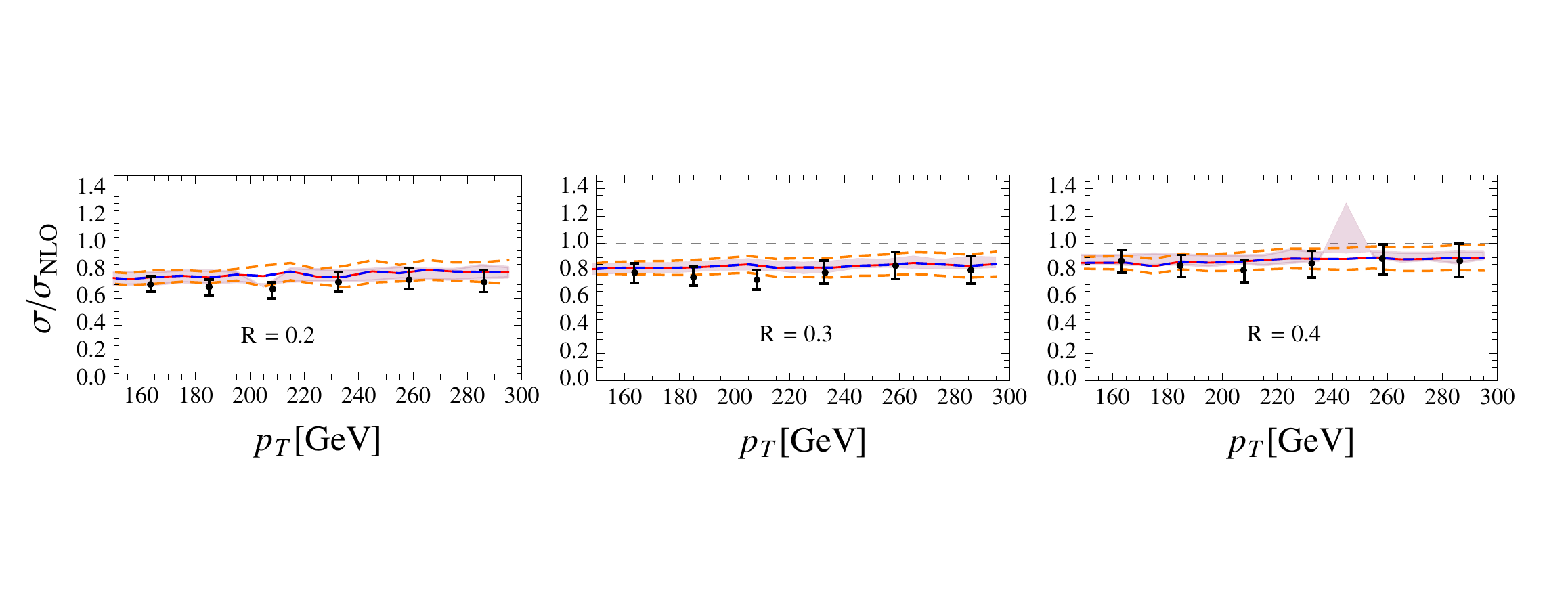,width=1.\textwidth,angle=0}
\end{center}
\vspace*{-2.2cm}
\caption{The resummed calculation for inclusive jet production for $|\eta|<2$ at $\sqrt{S} = 2.76$~TeV for different values of $R$ and the CMS data (black dots) of~\cite{Khachatryan:2016jfl} both normalized to the NLO results.
\label{fig2} }
\vspace*{-0.2cm}
\end{figure*}
%

We first present the results for the single-inclusive jet cross sections at $\sqrt{S} = 2.76$~TeV which was measured by the CMS Collaboration for different jet radii~\cite{Khachatryan:2016jfl}. In Fig.~\ref{fig2}, we show the resummed calculations using the CT10nlo PDFs~\cite{Lai:2010vv} for $|\eta| < 2$ along with the experimental data both normalized to the full NLO results. 
The error bars include both the experimental and the NLO scale
uncertainties added in quadrature as provided in~\cite{Khachatryan:2016jfl}. The dashed orange lines represent the PDF
uncertainty and the dashed blue lines show the PDF uncertainty for
the ratio $\sigma/\sigma_{\rm NLO}$ which is obtained by keeping the correlations. We
observe a significant improvement of the description of the data for all
values $R$ once the joint resummation is taken into account. 

\begin{figure}[t]
\begin{center}
\vspace*{-9mm}
\hspace*{-7mm}
\epsfig{figure=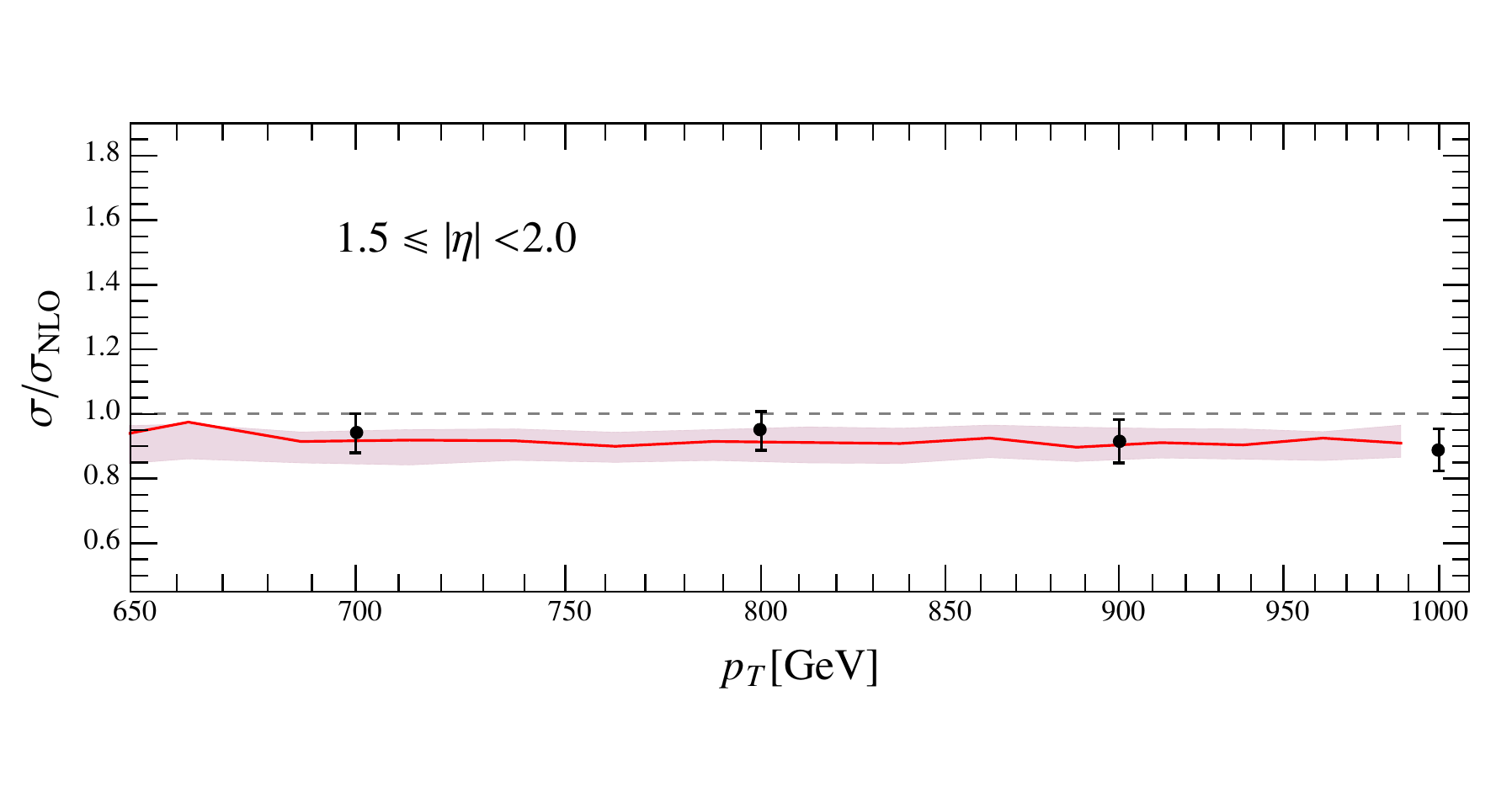,width=.54\textwidth,angle=0}
\end{center}
\vspace*{-12mm}
\caption{{The resummed calculation for inclusive jet production with $R=0.4$ at $\sqrt{S} = 13$~TeV and the preliminary ATLAS data (black dots) extracted from~\cite{ATLAS-CONF-2017-048} both normalized to the NLO result.
\label{fig3} }}
\vspace*{-.5cm}
\end{figure}

Next, we turn to the single-inclusive jet production at $\sqrt{S} = 13$~TeV.  The cross section was measured by ATLAS with a jet radius of $R=0.4$ for various bins of the jet-rapidity $\eta$~\cite{ATLAS-CONF-2017-048}. For the scale choice $\mu = p_T^{\rm max}$, the NLO predictions slightly overshoot the data by about $7\%$ to $10\%$ using the MMHT2014nlo PDF set for $|\eta|>1$. Nevertheless, the NLO calculation is still within the experimental errors bars. The NNLO corrections further enhance the cross section leading to a more significant disagreement with the data~\cite{ATLAS-CONF-2017-048}. In Fig.~\ref{fig3}, we show the results for the $p_T$ spectrum of our jointly resummed calculation. As an example, we consider the rapidity region $1.5 < |\eta| < 2$ and we plot the ratio of the NLL improved result to the NLO prediction. Here the error bars show both the NLO and PDF uncertainties~\cite{ATLAS-CONF-2017-048}. We find
that the joint resummation decreases the cross section relative to the NLO result and thus improves the agreement with the data. A similar trend can be observed when comparing to other single-inclusive jet analyses~\cite{Chatrchyan:2014gia,Khachatryan:2016wdh, Aaboud:2017dvo}. More detailed and systematic studies along those lines will be presented elsewhere~\cite{toappear}. 

{\it Conclusions.} In this work, we presented for the first time a joint resummation framework for single-inclusive jet production in the threshold and small-$R$ limit using SCET. Due to the small jet-size parameter used in the experimental analyses and the shape of the steeply falling luminosity functions, the threshold and the small-$R$ logarithmic terms make up the dominant bulk of the FO contributions in the kinematic range from moderate to large jet-$p_T$. Therefore, in order to provide reliable theoretical calculations, these classes of logarithmic corrections have to be resummed to all orders in perturbation theory. The fact that the full NNLO calculation depends significantly on the scale choice~\cite{ATLAS-CONF-2017-048} makes the importance of including higher-order corrections beyond NNLO even more evident. 

Using our framework, we obtained the resummed results for single-inclusive jet production at the LHC differential in both the jet-$p_T$ and the rapidity $\eta$. The scales in our framework are naturally chosen to minimize
  the logarithmic contributions arising due to the small values of $R$ and
  $z$. Instead, for the FO calculations there are no preferable scale choices that can avoid the
  occurrence of the large logarithms, although a lower scale choice may
  capture part of the resummation effects, similar to the choice $\mu_F = \mu_R =
  m_H/2$ in the case of Higgs-boson production in the gluon-gluon fusion channel~\cite{deFlorian:2014vta,Anastasiou:2016cez}.
We demonstrated the validity of our framework by finding very good agreement with the full NLO results for various values of $R$ and cuts on $p_T$ and $\eta$. We calculated all the necessary ingredients for the resummation to NLL accuracy and we presented phenomenological results for LHC kinematics at CM energies of $\sqrt{S} = 2.76$ and $13$~TeV. In both cases, we found improved agreements of the theoretical calculations with the LHC data after complementing the NLO calculations with the NLL resummation. We would like to stress again that the observed improvements are not limited to the exemplary data sets presented in this letter. Similar improvements are observed when comparing with other experimental analyses~\cite{Chatrchyan:2014gia,Khachatryan:2016wdh, Aaboud:2017dvo} at various machine energies covering a wide range of jet transverse momentum and rapidity, which will be presented in a forthcoming publication~\cite{toappear}.

The results presented in this work will have direct impacts on various aspects of QCD precision studies. 
On the phenomenology side, 
this includes the precise extraction of PDFs~\cite{Bonvini:2015ira} and the QCD strong coupling constant~\cite{Alekhin:2016evh} as well as the improvement of parton shower Monte Carlo event generators~\cite{Nagy:2016pwq}. 
  The joint resummation for single-inclusive jet production 
  supports ongoing efforts to include threshold resummation
  for the hard matrix elements in PDF fits and demonstrates the stability of the
  perturbative expansion. 
  On the theory side, our framework allows for realizing the joint
  resummation at NNLL accuracy in a straightforward manner as we have sketched in the text. 
  Subsequently, this can be matched with the known NNLO calculations for further improvement of the precision. 
  The full NNLO threshold expansion also captures all the leading
  contributions down to the $\delta(z)$ term. 
  These can be obtained within our framework and provide a partial cross check 
  of the NNLO results~\cite{Currie:2016bfm,Currie:2017ctp} 
  which have not been checked independently by a second calculation thus far.

{\it Acknowledgement.}  
We would like to thank Raghav Elayavalli, Jun Gao, Zhong-Bo Kang, Nobuo Sato, Ding Yu Shao and Werner Vogelsang for helpful communications. X.L. and S.M. were supported by the Munich Institute for Astro- and Particle Physics (MIAPP) of the DFG cluster of excellence "Origin and Structure of the Universe". X.L. would like to thank the Aspen Center for Physics for hospitality, which is supported by National Science Foundation grant PHY-1607611. S.M. acknowledges contract 05H15GUCC1 by BMBF. F.R. is supported by the U.S. Department of Energy under Contract No.~DE-AC02-05CH11231 and by the LDRD Program of Lawrence Berkeley National Laboratory. 

\bibliographystyle{h-physrev}
\bibliography{bibliography}

\end{document}